\documentclass[aip,pop,reprint,amsmath,amssymb]{revtex4-1}

\usepackage{graphicx}
\renewcommand{\vec}[1]{\boldsymbol{#1}}
\newcommand{\pp}[1]{\dot{#1}}
\newcommand{\ppp}[1]{\ddot{#1}}

\def\cpc{{Comput. Phys. Commun.} }
\def\pop{{Phys. Plasmas} }
\def\jcp{{J. Comput. Phys.} }
\def\jpp{{J. Plasma Phys.} }
\def\ppcf{{Plasma Phys. Control. Fusion} }
\def\pre{{Phys. Rev. E} }
\def\apjs{{Astrophys. J. Suppl. S.} }

\begin{document}

\title{On the Boris solver in particle-in-cell simulation}

\author{Seiji Zenitani}
\affiliation{Research Institute for Sustainable Humanosphere, Kyoto University, Gokasho, Uji, Kyoto 611-0011, Japan; Email: zenitani@rish.kyoto-u.ac.jp.}
\author{Takayuki Umeda}
\affiliation{Institute for Space-Earth Environmental Research, Nagoya University, Nagoya, Aichi 464-8601, Japan}

\begin{abstract}
A simple form of the Boris solver in particle-in-cell (PIC) simulation is proposed.
It employs an exact solution of the Lorentz-force part, and
it is equivalent to the Boris solver with a gyrophase correction. 
As a favorable property for stable schemes,
this form preserves a volume in the phase space. 
Numerical tests of the Boris solvers are conducted
by test-particle simulations and by PIC simulations.
The proposed form provides better accuracy than the popular form,
while it only requires few additional computation time.
\end{abstract}

\maketitle

\section{Introduction}
Particle-in-cell (PIC) method \citep{hockney,dawson83,birdsall,v05} is
one of the most important techniques in modern plasma simulations.
Solving Lagrangian motions of many charged particles
and the temporal evolution of Eulerian electromagnetic fields,
it can simulate various kinetic phenomena,
whose length scale is larger than the grid size. 

The particle integrator, an algorithm to advance charged particles,
is a fundamental element of PIC simulation. 
Since the particle integrator is used for all the particles,
its accuracy, stability, and computational cost
has a big impact on those of the entire simulation run.
One of the most common integrators is
the Boris solver \citep{boris70},
also known as the Buneman--Boris solver.
It solves the particle motion in a leap-frog manner, and
the acceleration part is split into several parts,
as will be shown later.
Owing to its simplicity and reliability,
the Boris solver has been used for nearly 50 years. 

In addition to the Boris solver,
various solvers have been developed.
For earlier ones, we refer the readers to
classic textbooks\citep{hockney,birdsall} and references therein.
Recently, particle solvers have been developed
much more actively than before.\citep{vay08,petri17,qiang17,umeda18,qin13,zhang15,HC17,ripperda18} 
\citet{vay08} has developed a solver
to carefully deal with the force balance.
By splitting the integrator into
the explicit first half and the implicit second half,
his solver better deals with
the {\bf E}$\times${\bf B} drift at a relativistic speed,
and therefore it has drawn growing attention
in laser physics and in astrophysics. 
\citet{petri17} has proposed a relativistic implicit solver,
which iteratively uses a matrix-form of the Vay solver for non-staggerd timesteps. 
\citet{qiang17} has presented a fast
Runge--Kutta relativistic integrator,
ready for the force balance problem.
\citet{umeda18} has proposed
a multi-step extension of the Boris integrator.
His solver effectively deals with the gyration
at a half timestep. 

From the theoretical viewpoint,
\citet{qin13} have recently pointed out that
the nonrelativistic Boris solver is stable, because
it preserves a volume in the phase space every timestep.
This property can be examined by a simple Jacobian matrix.
Presently, the volume preservation is regarded as
a key property for stable solvers.
\citet{zhang15} split the scheme
into several substeps
to discuss the volume preservation of their solver,
which appears to be another expression of the Boris solver.
\citet{HC17} have proposed a relativistic volume-preserving solver,
which employs Vay's characteristic velocity
in the Lorentz-force part of the Boris solver.
\citet{ripperda18} extensively compared
selected set of particle solvers.

In this contribution, we propose
a simple form of the Boris solver,
based on an exact solution for the Lorentz-force part.
We further examine the volume preservation of
the proposed form of the Boris solver.
Then we will present numerical tests of the Boris solvers,
followed by discussion and summary.

\section{Boris solver}
First, we outline the Boris algorithm.\citep{boris70} 
It handles the particle motion in the following discrete forms,
\begin{align}
\frac{\vec{x}^{n+\frac{1}{2}} - \vec{x}^{n-\frac{1}{2}}}{\Delta t} &= 
\frac{\vec{u}^{n}}
{\gamma^{n}}
\label{eq:x}
\\
m~\frac{\vec{u}^{n+1} - \vec{u}^{n}}{\Delta t} &= 
q \Big( \vec{E}^{n+\frac{1}{2}} + \vec{\bar{v}}^{n+\frac{1}{2}} \times \vec{B}^{n+\frac{1}{2}} \Big)
\label{eq:u}
\end{align}
where the superscripts ($n, n+1$, ...) indicate timesteps,
$\vec{u}=\gamma\vec{v}$ is the spatial part of a 4-vector,
$\gamma=[1-(v/c)^2]^{-1/2}=[1+(u/c)^2]^{1/2}$ is the Lorentz factor,
and $\vec{\bar{v}}$ is an effective velocity.
Other symbols have their standard meanings. 
The action part is split into
the Coulomb force for the first half timestep,
the Lorentz force for the entire timestep, and
the Coulomb force for the second half:
\begin{align}
\vec{u}^{-} &= \vec{u}^{n} + \vec{\varepsilon} \Delta t \label{eq:first} \\
\frac{\vec{u}^{+} - \vec{u}^{-}}{\Delta t} &= 
\frac{q}{m} \Big( \vec{\bar{v}}^{n+\frac{1}{2}} \times \vec{B}^{n+\frac{1}{2}} \Big)
\label{eq:second} \\
\vec{u}^{n+1} &= \vec{u}^{+} + \vec{\varepsilon} \Delta t
\label{eq:third}
\end{align}
where $\vec{\varepsilon} = ({q}/{2m}) \vec{E}^{n+\frac{1}{2}}$,
$\vec{u}^{-}$ and $\vec{u}^{+}$ are two intermediate states.
Hereafter we denote the field quantities $\vec{E}^{n+\frac{1}{2}}$, $\vec{B}^{n+\frac{1}{2}}$ at $n=n+\frac{1}{2}$ as $\vec{E}$, $\vec{B}$ for brevity.
Since Eq.~\eqref{eq:second} is
an energy-conserving rotation in the momentum space,
the Lorentz factor is set to be constant during the operation,
$\gamma^{-}=\gamma^{+}$.
The phase angle in the rotation part is
\begin{align}
\theta &= \frac{q\Delta t}{m \gamma^{-}} B
\label{eq:angle}
.
\end{align}
\begin{subequations}
The rotation is solved in the following way\\
\begin{minipage}{.42\columnwidth}
\begin{align}
\vec{t} = \tan\frac{\theta}{2}~\vec{b}
~~
\label{eq:boris00}
\end{align}
\end{minipage}
\begin{minipage}{.54\columnwidth}
\begin{align}
\vec{t} = \frac{\theta}{2}~ \vec{b} = \frac{q\Delta t}{2m \gamma^{-}} \vec{B}
~~
\label{eq:boris0} 
\end{align}
\end{minipage}
\end{subequations}

\begin{align}
\vec{u}' &= 
\vec{u}^{-} + \vec{u}^{-}\times \vec{t}
\label{eq:boris1} \\
\vec{u}^{+} &= \vec{u}^{-} + \frac{2}{ 1 + t^2 }
\Big(
\vec{u}'\times \vec{t}
\Big)
\label{eq:boris2} .
\end{align}
where $\vec{b} = {\vec{B}}/|B|$ is a unit vector.
There are two choices in Eq.~(7).
One can advance $\vec{u}^{n}$ to $\vec{u}^{n+1}$ by using
either of the two equation sets.
\citet{boris70} presented
a procedure of Eqs.~\eqref{eq:first}, \eqref{eq:angle}, \eqref{eq:boris00}, \eqref{eq:boris1}, \eqref{eq:boris2}, and \eqref{eq:third}
in his original article. 
The subsequent textbooks\citep{hockney,birdsall} described
a simplified procedure [Eqs.~\eqref{eq:first}, \eqref{eq:boris0}, \eqref{eq:boris1}, \eqref{eq:boris2}, and \eqref{eq:third}],
by replacing Eq.~\eqref{eq:boris00} with Eq.~\eqref{eq:boris0}.
We call the two procedures the Boris-A solver and the Boris-B solver, respectively.

\begin{figure}[htbp]
\includegraphics[width={.8\columnwidth}]{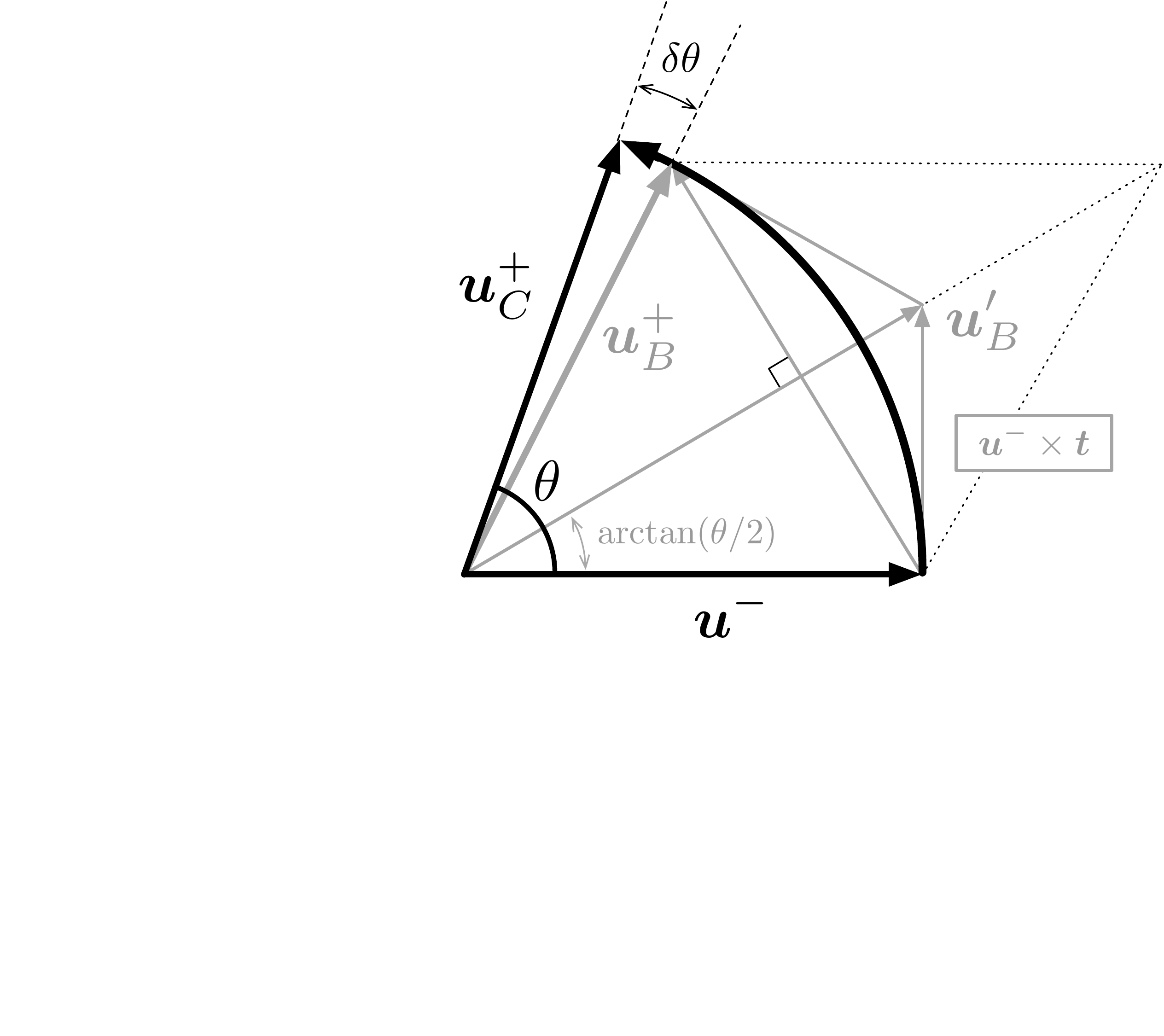}
\caption{
Schematic diagram for the Lorentz-force part of the Boris solvers.
Procedures by the Boris-C solver (in black) and by the Boris-B solver (in gray) are illustrated.
There is an error in the phase angle ($\delta\theta$) in the Boris-B solver.
\label{fig:scheme}}
\end{figure}

Although we do not describe the detail, 
the Boris-A solver accurately handles the rotation. 
As a result of the replacement (Eq.~\eqref{eq:boris00} $\rightarrow$ Eq.~\eqref{eq:boris0}),
the Boris-B solver approximates the rotation,
as illustrated in gray in Figure \ref{fig:scheme}. 
There is always a delay in the gyrophase angle,
\citep{birdsall}
\begin{align}
{\delta \theta} = \theta - 2\arctan\frac{\theta}{2}
= \theta \Big( \frac{1}{12}\theta^2- \frac{1}{80}\theta^4+\cdots \Big)
.\label{eq:phase}
\end{align}
Thus, the Boris-B solver is an approximate form of the original Boris solver (the Boris-A solver).
Despite this, owing to its simplicity and computational cost,
the Boris-B solver is widely used.
Scientists often indicate the Boris-B solver
simply by ``the Boris solver.''
Eqs.~\eqref{eq:boris00} and \eqref{eq:boris0} differ
by a factor of $\tan (\frac{\theta}{2})/ (\frac{\theta}{2})$.
Therefore, the Boris-A solver is sometimes referred to as
the Boris solver with a gyrophase correction factor.\citep{hockney,birdsall}

\section{Exact gyration solver}
\label{sec:solver}
To avoid the phase error in the Boris-B solver,
one can use the Boris-A solver. 
Here, we propose another expression.
We propose the following rotation procedure based on an analytic solution:
\begin{align}
\label{eq:new1}
\vec{u}^{-}_{\parallel} &= {(\vec{u}^{-}\cdot\vec{b})~\vec{b}} \\
\label{eq:new2}
\vec{u}^{+} &= \vec{u}^{-}_{\parallel} + (\vec{u}^{-}-\vec{u}^{-}_{\parallel}) \cos \theta + ({\vec{u}^{-} \times \vec{b}}) \sin \theta
\end{align}
This is illustrated in black in Figure \ref{fig:scheme}. 
Then, one can advance the particle by
Eqs.~\eqref{eq:first}, \eqref{eq:angle}, \eqref{eq:new1}, \eqref{eq:new2},
and \eqref{eq:third}
in a time-reversible manner. 
We call this the Boris-C solver.
In practice, we use a threshold $\epsilon_b$ in $|B|^2 = \max({\vec{B}^2, \epsilon_b})$, to avoid dividing by zero.
This does not cause a significant error in the $B\rightarrow 0$ limit.

The Boris-C solver provides second-order accuracy.
The splitting into the Coulomb-force part (Eqs.~\eqref{eq:first} and \eqref{eq:third}) and
the Lorentz-force part (Eqs.~\eqref{eq:new1}--\eqref{eq:new2}) is equivalent to an operator splitting,
also known as the Strang splitting.\cite{strang68,laveque02}
The Strang splitting gives second-order accuracy,
if each operator maintains more than second-order accuracy.
In this case, since both of the two parts give exact solutions,
the resulting scheme should have second-order accuracy.

One can also prove the second-order accuracy in a straightforward manner. 
In the constant fields,
the second-order expansion of
the new state $\vec{u}^{n+1}$ is
\begin{align}
\vec{u}^{n+1} = \vec{u}^{n} + \pp{\vec{u}}^{n} \Delta t + \frac{1}{2} \ppp{\vec{u}}^{n}\Delta t^2 + \mathcal{O}(\Delta t^3)
\label{eq:order}
\end{align}
where
\begin{align}
\pp{\vec{u}} &= \frac{q}{m}( \vec{E} + \vec{v}\times\vec{B} )
\label{eq:order1} \\
\ppp{\vec{u}}
&= \frac{q}{m}( \pp{\vec{v}}\times\vec{B} )
= \frac{q}{m}
\Big(
\frac{\pp{\vec{u}}}{\gamma}
-
\frac{\vec{u}\cdot\pp{\vec{u}}}{\gamma^3c^2}\vec{u}
\Big)
\times\vec{B}
\nonumber \\
&= \Big(\frac{q}{m}\Big)^2
\Big(
\frac{\vec{E}+\vec{v}\times\vec{B}}{\gamma}
-
\frac{\vec{u}\cdot\vec{E}}{\gamma^3c^2}\vec{u}
\Big)
\times\vec{B}
\label{eq:order2}
\end{align}
Following the Boris procedure, we find
\begin{align}
\vec{u}^{-} &= \vec{u}^{n} + \vec{\varepsilon}\Delta t
\label{eq:borisorder1} \\
\vec{u}^{+} &= \vec{u}^{-} + \frac{q}{m}( \vec{v}^{-}\times\vec{B} ) \Delta t
\nonumber \\
&~~ + \frac{1}{2} 
\Big\{
\Big(\frac{q}{m}\Big)^2
\Big(
\frac{\vec{v}^{-}\times\vec{B}}{\gamma^{-}}
\Big)
\times\vec{B}
\Big\} \Delta t^2
+ \mathcal{O}(\Delta t^3) + \cdots
\label{eq:borisorder2} \\
\vec{u}^{n+1}
&= \vec{u}^{+} + \vec{\varepsilon}\Delta t
\label{eq:borisorder3}
\end{align}
In Eq.~\eqref{eq:borisorder2},
the Lorentz-force part is expanded for $\vec{u}^{-}$, similarly as Eqs.~\eqref{eq:order}--\eqref{eq:order2},
but with $\vec{E}=0$.
Although the Boris-C solver accurately solves
the third and higher-order terms in Eq.~\eqref{eq:borisorder2},
we focuses on the terms up to second-order.
Importantly, Eqs.~\eqref{eq:borisorder1} and \eqref{eq:borisorder3} do not contain second or higher-order terms, $\mathcal{O}(\Delta t^2)$, because the solver already gives an exact solution for the Coulomb-force part.
From Eq.~\eqref{eq:borisorder1},
we obtain first-order expansions of
the following variables,
\begin{align}
\vec{v}^{-}
&= \frac{1}{\gamma}
\Big( \vec{u}^{n} + \frac{q}{2m}\vec{E}\Delta t \Big)
\Big( 1+ \frac{q\vec{u}^{n}\cdot\vec{E}}{mc^2\gamma^2}\Delta t + \mathcal{O}(\Delta t^2)\Big)^{-1/2} \nonumber \\
&=
\vec{v}^{n} +
\frac{q}{2m\gamma} \Big( \vec{E}
- \frac{\vec{u}^{n}\cdot\vec{E}}{c^2\gamma^2} \vec{u}^{n} \Big) \Delta t + \mathcal{O}(\Delta t^2)
\label{eq:vorder1} \\
\frac{\vec{v}^{-}}{\gamma^{-}}
&=
\frac{\vec{v}^{n}}{\gamma} +
\frac{q}{2m\gamma} \Big( \vec{E}
- \frac{2\vec{u}^{n}\cdot\vec{E}}{c^2\gamma^2} \vec{u}^{n} \Big) \Delta t + \mathcal{O}(\Delta t^2)
\label{eq:vorder2}
\end{align}
Substituting Eqs.~\eqref{eq:vorder1}--\eqref{eq:vorder2} into
Eqs.~\eqref{eq:borisorder1}--\eqref{eq:borisorder3},
one can obtain an expanded form of the Boris-C solver,
$\vec{u}^{n+1} = \vec{u}^{n} + \cdots +  \mathcal{O}(\Delta t^3)$.
The first and second-order coefficients are identical to those of
the Talyor expansion (Eq.~\eqref{eq:order}--\eqref{eq:order2}) for $\vec{u}=\vec{u}^{n}$.
This proves that
the Boris-C solver has second-order accuracy.


%
%
As a theoretical property,
we examine the volume preservation\citep{qin13} of the Boris-C solver. 
This discussion holds true for the Boris-A solver as well,
because both methods accurately solve the rotation.
We consider the temporal evolution of a phase-space volume,
by splitting the Boris-C solver to the following substeps.\citep{zhang15}
We consider half-adjusted positions $x^{n}$ and $x^{n+1}$.
$R$ stands for a 3-D rotation matrix.
\begin{align}
&1: &
\vec{x}^{n+\frac{1}{2}} &=
\vec{x}^{n} + \frac{\Delta t}{2} \frac{\vec{u}^{n}}{\gamma^{n}},
&
\vec{u}^{n} &= \vec{u}^{n}
\label{eq:VPA1}\\
&2: &
\vec{x}^{n+\frac{1}{2}} &= \vec{x}^{n+\frac{1}{2}},
&
\vec{u}^{-} &= \vec{u}^{n} + \vec{\varepsilon} \Delta t
\label{eq:VPA2}\\
&3: &
\vec{x}^{n+\frac{1}{2}} &= \vec{x}^{n+\frac{1}{2}},
&
\vec{u}^{+} &= R \vec{u}^{-}
\label{eq:VPA3}\\
&4: &
\vec{x}^{n+\frac{1}{2}} &= \vec{x}^{n+\frac{1}{2}},
&
\vec{u}^{n+1} &= \vec{u}^{+} + \vec{\varepsilon} \Delta t
\label{eq:VPA4}\\
&5: &
\vec{x}^{n+1} &=
\vec{x}^{n+\frac{1}{2}} + \frac{\Delta t}{2} \frac{\vec{u}^{n+1}}{\gamma^{n+1}},
&
\vec{u}^{n+1} &= \vec{u}^{n+1}
\label{eq:VPA5}
\end{align}
We evaluate the Jacobian $J_k$ for the $k$-th substep,
\begin{align}
J_k
=
\Big| \frac{\partial(\vec{x}^{new},\vec{u}^{new})}{\partial(\vec{x}^{old},\vec{u}^{old})} \Big|
=
{\rm det}
\begin{pmatrix}
\cfrac{\partial \vec{x}^{new}}{\partial \vec{x}^{old}} &
\cfrac{\partial \vec{x}^{new}}{\partial \vec{u}^{old}} \\
\cfrac{\partial \vec{u}^{new}}{\partial \vec{x}^{old}} &
\cfrac{\partial \vec{u}^{new}}{\partial \vec{u}^{old}} \\
\end{pmatrix}
.
\end{align}
From Eqs.~\eqref{eq:VPA1}--\eqref{eq:VPA5},
we obtain
\begin{align}
&
J_1 = J_5 =
\begin{vmatrix}
{I} &
\frac{\partial \vec{x}^{new}}{\partial \vec{u}^{old}} \\
0 & {I} \\
\end{vmatrix}
= 1,
~~
J_2 = J_4 =
\begin{vmatrix}
{I} & 0 \\
\frac{\partial \vec{u}^{new}}{\partial \vec{x}^{old}}
& {I} \\
\end{vmatrix}
= 1,\\
&
J_3 =
\begin{vmatrix}
{I} & 0 \\
\frac{\partial \vec{u}^{new}}{\partial \vec{x}^{old}}
&
\frac{\partial \vec{u}^{new}}{\partial \vec{u}^{old}}
\\
\end{vmatrix}
= \Big| \frac{\partial \vec{u}^{new}}{\partial \vec{u}^{old}} \Big|.
\label{eq:J3}
\end{align}
The third step (Eqs.~\eqref{eq:VPA3} and \eqref{eq:J3}) corresponds
to a $\gamma$-dependent rotation, i.e., $\theta \propto \gamma^{-1}$. 
This case, we confirm
$J_3=|{\partial (\vec{u}^{new})}/{\partial (\vec{u}^{old})}| = 1$
as shown in Appendix. 
Thus, we obtain $J_k=1$ for $k=1 \cdots 5$.
This indicates that
the solver preserves a volume in the phase space every substep.
Therefore, the Boris-C solver is volume-preserving
during the entire step from
$(\vec{x}^{n}, \vec{v}^{n})$ to $(\vec{x}^{n+1}, \vec{v}^{n+1})$.

For comparison, we examine the volume preservation
for the nonrelativistic fourth-order Runge--Kutta method in Appendix \ref{sec:RK}.
The phase-space volume is not preserved in this case,
$J_{\rm RK4} \ne 1$.

\begin{table}
\begin{tabular}{llll}
\hline
\#
~~
 & 
$\vec{B}$~~~~~~~~ &
$\vec{E}$~~~~~~~~ &
~\\
\hline
1 &
(0,0,0) & (1,0,0) 	& {\rm direct acceleration by {\bf E}} \\
2 &
(0,0,0.1) & (1,0,0) 	& {\rm {\bf E}-dominated} \\
3 &
(0,0,1) & (1,0,0) 	& {\rm $|E|=|B|$} \\
4 &
(0,0,1) & (0.1,0,0) & {\rm {\bf E}$\times${\bf B} drift} \\
5 &
(0,0,1) & (0,0,0) 	& {\rm gyration about {\bf B}} \\
\hline
\end{tabular}
\caption{
Field settings for test-particle simulations.
\label{table}}
\end{table}

\begin{figure*}[htbp]
\includegraphics[width={\textwidth}]{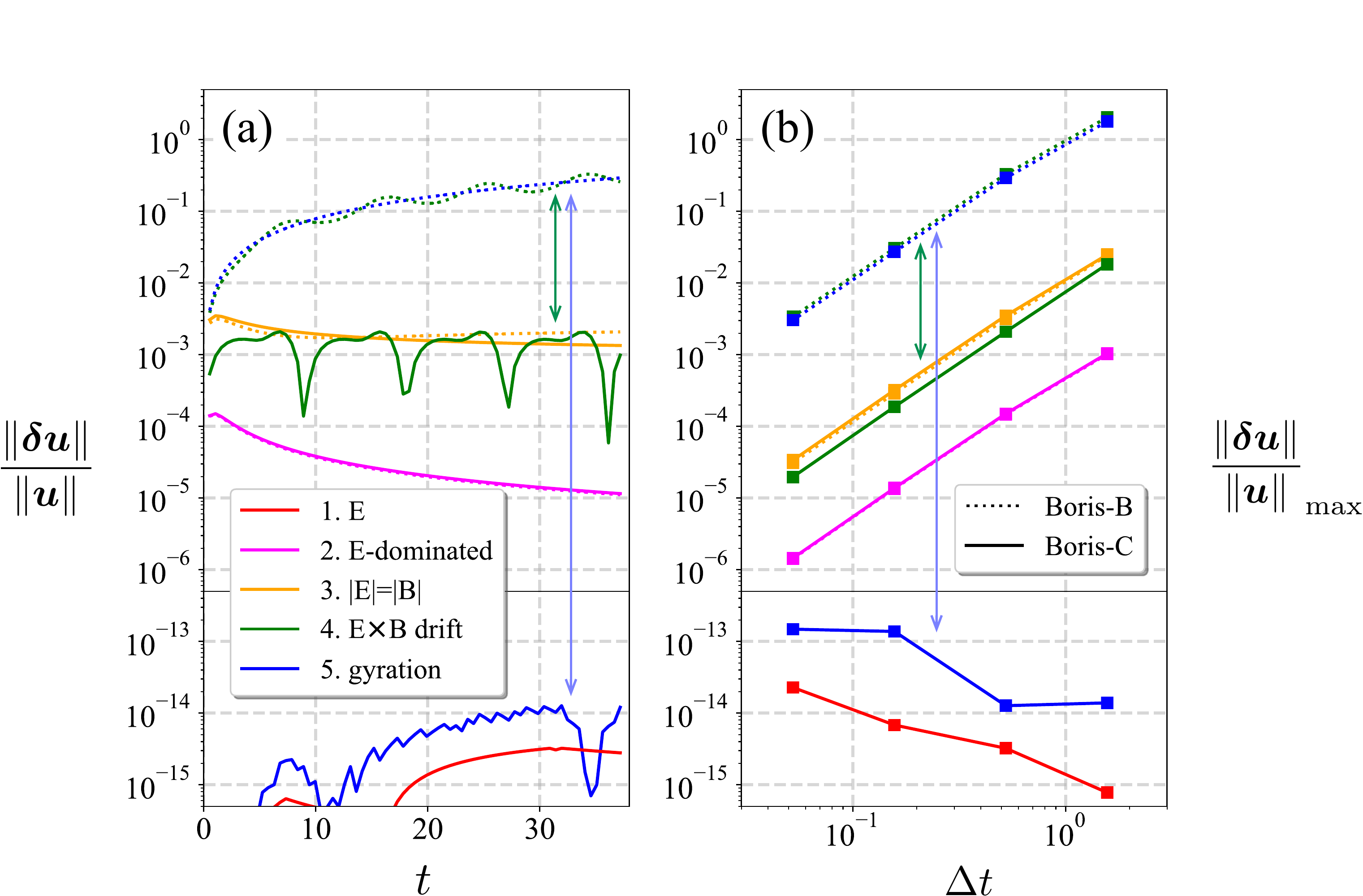}
\caption{
(a) Time evolution of numerical errors $|\delta\vec{u}|/|{u}|$ in test-particle simulations.
The dotted lines indicate the results by the Boris-B solver,
whereas the solid lines indicate the results by the Boris-C solver.
The timestep is fixed to $\Delta t= \pi /6$. 
(b) Maximum errors, $(|\delta\vec{u}|/|{u}|)_{\rm max}$ during $0<t<12\pi$ as a function of $\Delta t$.
\label{fig:error}}
\end{figure*}

\section{Numerical tests} 
In order to see the accuracy, stability, and performance of the Boris solvers, we have carried out four numerical tests. 
First, to check the accuracy,
we have carried out test-particle simulations
in a uniform electromagnetic field.
Physics parameters are set to $m=1$, $q=1$, $c=1$, and
$\vec{u}(t=0)=(1,0,0)$.
The time interval, the timestep, and the numerical threshold are
set to $0<t<12/\pi$, $\Delta t = \pi/6$, and $\epsilon_b = 10^{-20}$.
We consider five configurations,
as shown in Table \ref{table}.
The first case corresponds to direct acceleration by the electric field.
A weak magnetic field is imposed in the second case, but
the electric field is still dominant.
The third is a special case of $\vec{E}\perp\vec{B}$ and $|{E}|=|{B}|$. 
The fourth case corresponds to the {\bf E}$\times${\bf B} drift.
The drift is slightly modulated by the relativistic effect.
The last case corresponds to gyration about the magnetic field. 
In each cases, we evaluate
an error in 4-vector to a reference value,
$\delta\vec{u}=\vec{u}-\vec{u}_{\rm ref}$. 
For cases 1 and 5, we employ analytic solutions as reference values.
For cases 2--4, we refer numerical results
with a small timestep $\Delta t = \pi/240$,
because we do not know analytic solutions
as a simple function of $t$. 
The numerical reference values are checked
by analytic solutions in other forms.\citep{book2,friedman05}

Figure \ref{fig:error}(a) shows
temporal evolution of errors
$|\delta\vec{u}|/|{u}|$ in our test-particle simulations.
The dotted lines represent results by the Boris-B solver,
whereas the solid lines represent results by the Boris-C solver.
The results by the Boris-A solver are not shown,
because they are essentially the same as those by the Boris-C solver.
In case 1 (in red),
the Boris-B and Boris-C solvers use the same parts (Eqs.~\eqref{eq:first} and \eqref{eq:third}) and 
their results are identical. 
The two solvers give accurate results,
because Eqs.~\eqref{eq:first} and \eqref{eq:third}
give an exact solution for the linear acceleration by {\bf E}.
In cases 2 and 3 (in magenta and in orange),
the two solvers give similar results. 
The curves drop,
because $|\delta\vec{u}|$ does not grow and
because $|{u}|$ increases in time. 
%
In case 4 of the {\bf E}$\times${\bf B} drift (in green),
the Boris-C solver drastically improves the results. 
It reduces the error by two orders-of-magnitude,
as indicated by the green arrow. 
The error by the Boris-B solver grows in time,
because the phase error
$|\delta\vec{u}| \approx |{u}|\delta \theta = \delta \theta$
accumulates in time. 
On the other hand, the error by the Boris-C solver remains small. 
The repeated drop corresponds to the gyroperiod. 
Since the Boris-C solver exactly solves the phase,
the solid curve repeatedly show the same pattern,
and it does not grow further. 
In case 5 of gyration (in blue),
the error by the Boris-B solver linearly grows in time ($\propto t$)
because of the accumulation of the phase error. 
In contrast, the Boris-C solver gives accurate results, as we have expected.

Figure \ref{fig:error}(b) presents
maximum errors $(| \delta\vec{u} | / | {u} |)_{\rm max}$ during $0<t<12\pi$
as a function of the timestep,
$\Delta t = \pi/60$, $\pi/20$, $\pi/6$, and $\pi/2$.
The results demonstrate the second-order accuracy
of the solvers. 
As evident in cases 1--3,
the Boris-C solver is as good as
the Boris-B solver when $|E| \gtrsim |B|$.
As the electric field dominates,
the errors decrease, and then
they give accurate results in case 1 (red line).
When $|B| \gtrsim |E|$,
the two solvers give different results.
In cases 4 and 5,
the Boris-B solver remains moderately good
(the green and blue dotted lines).
From Eq.~\eqref{eq:phase} and using $\gamma=\sqrt{2}$,
one can estimate the curve in case 5,
$|\delta \vec{u}|/|{u}| \approx (12\pi / \Delta t) \delta \theta = (\pi/\sqrt{8})\Delta t^2$,
in agreement with the blue dotted line. 
In contrast, the Boris-C solver produces drastically smaller errors than the Boris-B solver,
even though it maintains the second-order accuracy.
As the magnetic field dominates, the errors decrease, and then
the solver gives accurate results in case 5 (blue solid line).
In the bottom part in Figure \ref{fig:error}(b),
the errors in the two exact cases
probably come from a machine error.
For example, for case 5,
if an error of $\mathcal{O}(10^{-15})$ in double precision floating numbers is accumulated every timestep, 
the total error would be
$|\delta \vec{u}|/|{u}| \lesssim 10^{-15} \times (12 \pi / \Delta t) \approx 10^{-13.5} \Delta t^{-1}$,
in consistent with the blue line.
The error in case 1 (red line) is even smaller,
because of larger $|{u}|$.

\begin{figure}[htbp]
\includegraphics[width={\columnwidth}]{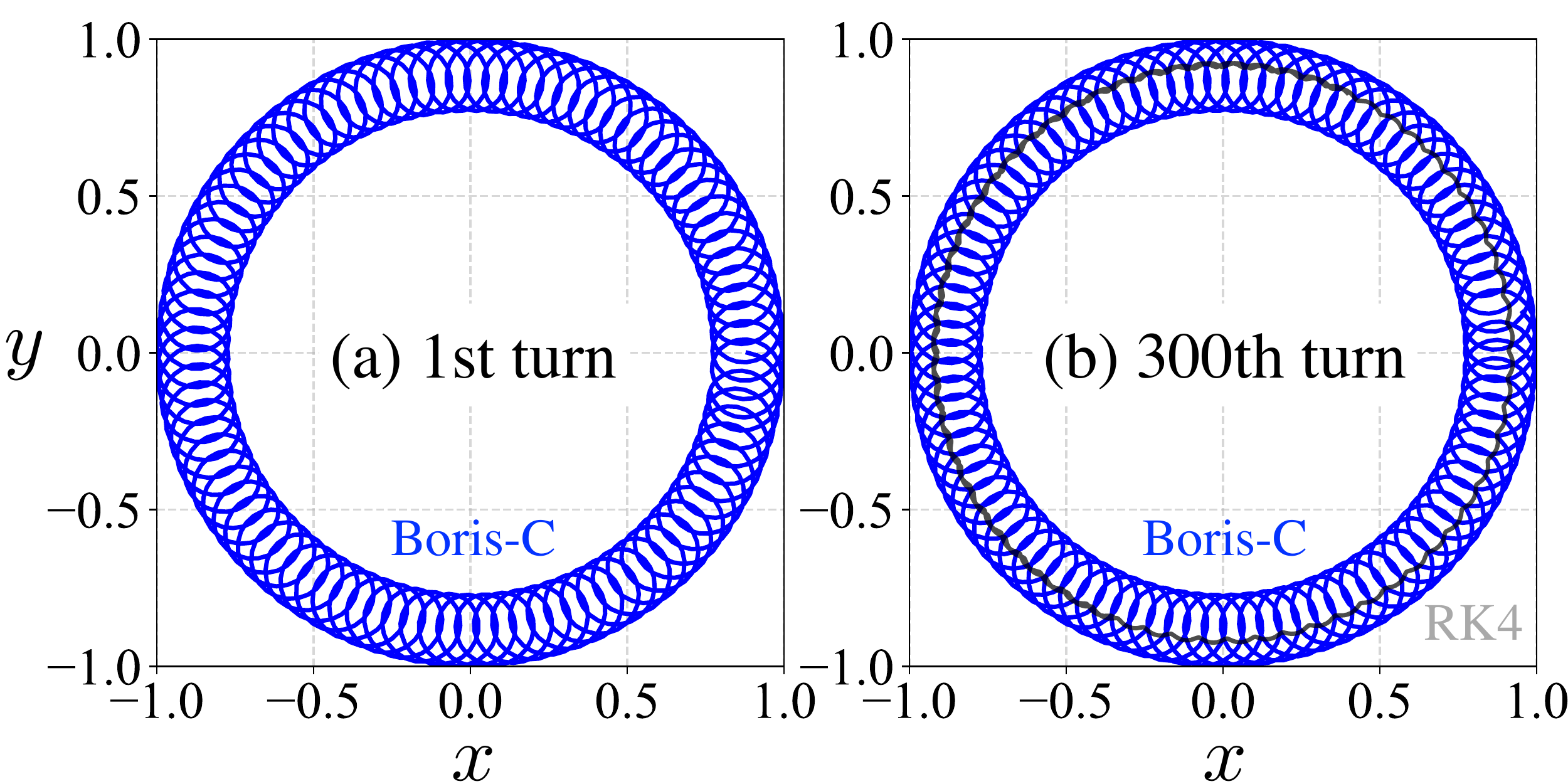}
\caption{
Particle trajectories by the Boris-C solver
during (a) the initial stage and (b) the late stage.
A trajectory by the RK4 solver is overplotted in gray in (b) the late stage.
\label{fig:qin}}
\end{figure}

As a second numerical test, 
we evaluate the long-term stability of the particle solvers.
We have run a code in the following field,
as was done in Ref.~\onlinecite{qin13},
\begin{align}
\vec{B} = (x^2+y^2)^{1/2} \vec{e}_z,
~~
\phi = 0.01 (x^2+y^2)^{-1/2}
.
\end{align}
The initial conditions are $\vec{u}(t=0)=(0.1,0.0)$,
$\vec{x}(t=0)=(0.9,0,0)$, $m=c=1$, and $\Delta t = \pi/10$.
Figure \ref{fig:qin} shows the trajectories
at (a) an initial stage and at (b) a late stage by the Boris-C solver (in blue)
and by the fourth order Runge--Kutta solver (in gray).
From the beginning, the particle undergoes
a fast small-scale gyration and
a slow large-scale rotation due
to the $\nabla B$ drift and the {\bf E}$\times{\bf B}$ drift. 
The large-scale drifts keep the particle
in the same domain, and then
we inspect the evolution of the trajectory. 
After a long time (300th turn; $t \approx 2\times 10^5$),
as evident in Fig.~\ref{fig:qin}(b),
the Runge--Kutta solver dissipates
the small-scale motion and the relevant kinetic energy.
This is because the Runge--Kutta solver is not volume-preserving,
at least in the uniform fields in the nonrelativistic regime (Appendix \ref{sec:RK}), and quite likely so in generic relativistic cases.
In contrast, the Boris-C solver is free from the numerical damping,
similarly as the Boris-B solver (not shown).
This numerical experiment demonstrates that
the Boris-C solver does preserve the phase-space volume for the small-scale gyration over a long time. 

As a third numerical test,
to compare computational costs,
we have run the case 5 in Table \ref{table}
until $t=2\times 10^7 \pi$ ($1.2 \times 10^8$ steps),
by only advancing the particle velocity. 
Average elapse times are proportional to $1.46:1:1.24$ for the Boris-A, B, and C solvers on our PC (Intel processor) and
$1.46:1:1.26$ on FX100 computer (SPARC processor)
at Japan Aerospace Exploration Agency (JAXA). 
While performance depends on
implementation, compilers, and CPU architectures,
the Boris-C solver runs faster than the Boris-A solver, and
it runs only 25\% slower than the Boris-B solver.

As a final test,
we have carried out PIC simulations of
a physics problem on magnetic reconnection. 
The settings are documented in Ref.~\onlinecite{zeni18},
and therefore we do not repeat them here. 
Comparing average elapse times,
we have found that
the Boris-C solver is slower than the Boris-B solver by 1.2\%
on XC50 supercomputer (Intel processor)
at National Astronomical Observatory of Japan and
by 3.3\% on FX100 computer (SPARC processor) at JAXA.

\section{Discussion and Summary}
As already discussed,
the Boris-C solver is a combination of the two exact solvers
for the Coulomb-force part and for the Lorentz-force part,
while the Boris-B solver is based on the exact Coulomb-force solver 
and a second-order solver for the Lorentz-force part.
From the viewpoint of operator-splitting, in both cases,
the combined solver maintains second-order accuracy, 
because each part has second-order accuracy.
This is evident in Figure \ref{fig:error}(b).
Since the Boris-C solver better deals with the Lorentz-force part,
the amplitude of the second-order error is much smaller than
in the Boris-B solver, in particular in the $|B| \gtrsim |E|$ cases. 
The Boris-B solver gives a second-order error in phase (Eq.~\eqref{eq:phase}). 
Other second-order solvers, such as the Higuera--Cary solver and the Umeda solver, could suppress the total error,
by better solving the phase.\citep{HC17,umeda18,ripperda18}
In this line, the Boris-C and A solvers provide the best possible results,
because they give no phase error in phase.

We have formally proved that
the Boris-C solver preserves a volume in the phase space
during the temporal evolution,
and then we have confirmed that
it preserves the small-scale gyration after a long-time computation. 
The volume preservation of the popular Boris-B solver
has been extensively studied,\citep{qin13,zhang15,HC17}
however, it has never been evaluated in the Boris-A or C solvers.
Since the Boris-C solver is simple,
the proof is given straightforwardly. 
This gives further confidence to
the reliablity of the Boris-C and A solvers. 

Many scientists prefer the Boris-B solver to
the Boris-A solver because of the low computation cost. 
In fact, according to the numerical test,
the Boris-A solver is 46\% more expensive than the Boris-B solver.
Since the Boris-C solver employs simple expressions,
it is more favorable than the Boris-A solver.
The Boris-C solver gives 25\% in test-particle simulations.
Obviously the conventional solvers (the Boris-A and Boris-B solvers)
tried to avoid trigonometric functions which were expensive at that time. 
Presently, these functions are not so expensive as they used to be,
and therefore the Boris-C solver runs adequately fast. 
In PIC simulations,
the Boris-C solver runs slower only by 1--3\% in PIC simulations. 
Typically, the particle integrator and the electric current calculator
account for the most of the computation time, and
they consume equal amount of time.
Then largely due to memory-access performance,
the particle integrator does not run at full speed. 
If we assume that it runs at 50\% efficiency,
the particle integrator is responsible for
25\% of the computation time.
Then, the 25\% slow-down in the particle integrator
should result in 6\% slow-down in the entire run. 
We have observed 1--3\% in our PIC simulations,
probably because our code runs less efficiently,
but the results are reasonable.
The computational cost can be easily compensated
by employing a larger $\Delta t$.

From the viewpoint of the stability, 
we usually keep the timestep small,
$\omega_{\rm max} \Delta t < 2$,
where $\omega_{\rm max}$ is the maximum frequency to resolve.\citep{birdsall}
Nevertheless, in PIC simulation,
the magnitude of the magnetic field
may instantly approach or exceed
the $\frac{qB}{\gamma mc} \Delta t = 2$ criteria, and so
it is useful to check the stability for a large $\Delta t$ limit. 
The Boris-B solver delays the gyrophase (Eq.~\eqref{eq:phase}),
and then the angle never exceed $\pi$, i.e., $\theta<\pi$.
The particle eventually moves back-and-forth.\citep{parker91}
The Boris-A and Boris-C solvers have no such limitation.
They simply allow particles to gyrate more than $\pi$,
although it leads to an opposite gyration. 
Meanwhile, the Boris-A solver needs some care for $\theta \approx \pi$,
where Eq.~\eqref{eq:boris00} is undefined or approaches $\pm\infty$. 
Therefore the Boris-B and Boris-C solvers are safer choices
among the three.

Finally, we note that there has been continual progress in developing relativistic symplectic solvers.\citep{wu03,shadwick14,zhang18,wolski18}
In fact, the symplectic algorithms\citep{channel90,yoshida93,marsden01} are favored in many applications, owing to their long-term accuracy. 
However, the symplectic schemes are often implicit and computationally expensive, while the number of explicit symplectic solvers for relativistic charged particles in arbitrary electromagnetic fields is limited.\citep{zhang18,wolski18}
In addition, it was reported that
volume-preserving solvers are sometimes more accurate than symplectic solvers.\citep{chin08}
Considering these issues, it will take some time before symplectic solvers will be popular in relativistic PIC simulations. 
Meantime, it is important to
improve the second-order Boris solvers,
which have an advantage in computational efficiency and
which are proven to be reliable.

In summary,
we have proposed a simple form of the Boris solver,
based on the exact solution for the Lorentz-force part.
It is equivalent to the Boris solver with a gyrophase correction.
It has a favorable property of
preserving the phase-space volume, and
therefore it appears to be stable.
The proposed form gives much more accurate results
than the popular form (the Boris-B solver),
while it only requires few additional computation time. 
We hope that the proposed form will be
a good alternative to the conventional Boris solvers.

\begin{acknowledgements}
The authors acknowledge
the anonymous reviewer for his/her professional remarks.
One of the author (SZ) acknowledges
T.-N. Kato, Y. Omura, and T. Nogi for useful comments.
This work was supported by
Grant-in-Aid for Scientific Research (C) 17K05673 and (S) 17H06140
from the Japan Society for the Promotion of Science (JSPS).
\end{acknowledgements}

\appendix

\section{Jacobian for relativistic gyration}
Without losing generality, 
we consider a gyration about $\vec{B}=(0,0,B)$.
Then the rotation procedure is expressed by
\begin{align}
\vec{u}^{new} = R \vec{u} =
\begin{pmatrix}
u_x \cos \frac{\theta_0}{\gamma} + u_y \sin \frac{\theta_0}{\gamma} \\
-u_x \sin \frac{\theta_0}{\gamma} + u_y \cos \frac{\theta_0}{\gamma} \\
u_z
\end{pmatrix}
\end{align}
where $\theta_0=(qB\Delta t)/m$.
Considering
\begin{align}
\frac{\partial}{\partial u_x} \Big( \frac{\theta_0}{\gamma} \Big)
=
\theta_0 \frac{\partial}{\partial u_x} \Big( 1+(u/c)^2 \Big)^{-\frac{1}{2}}
=
-\frac{\theta_0}{\gamma^3c^2} u_x
,
\end{align}
we obtain a Jacobian matrix
for $\vec{u} \rightarrow \vec{u}^{new}$:
\begin{align}
\frac{\partial (\vec{u}^{new})}{\partial (\vec{u})}
=
\begin{pmatrix}
~~
\cos \frac{\theta_0}{\gamma} - f_0 f_1 u_x&
\sin \frac{\theta_0}{\gamma} - f_0 f_1 u_y&
- f_0 f_1 u_z \\
-\sin \frac{\theta_0}{\gamma} - f_0 f_2 u_x &
 \cos \frac{\theta_0}{\gamma} - f_0 f_2 u_y &
- f_0 f_2 u_z \\
0 & 0 & 1
\end{pmatrix}
,
\label{eq:shear_rot1}
\end{align}
where
\begin{eqnarray}
\label{eq:spherical_scattering}
\left\{ \begin{array}{lll}
f_0 &=& \frac{\theta_0}{\gamma^3c^2}, \\
f_1 &=& -u_x \sin \frac{\theta_0}{\gamma} + u_y \cos \frac{\theta_0}{\gamma},\\
f_2 &=& -u_x \cos \frac{\theta_0}{\gamma} - u_y \sin \frac{\theta_0}{\gamma}.
\end{array} \right.
\label{eq:shear_rot2}
\end{eqnarray}
From Eqs.~\eqref{eq:shear_rot1} and \eqref{eq:shear_rot2},
some algebra leads to
\begin{align}
\Big| \frac{\partial (\vec{u}^{new})}{\partial (\vec{u})} \Big|
=& 1.
\end{align}

\section{Volume preservation for Runge--Kutta method}
\label{sec:RK}

Here we focus on a simple case of
nonrelativistic particle motion in a uniform magnetic field.
The fourth-order Runge--Kutta method advance a particle in the following way:
\begin{align}
\vec{v}^{n+1} =&~ \vec{v}_n + \frac{1}{6} (\vec{k}_1+2\vec{k}_2+2\vec{k}_3+\vec{k}_4)\Delta t \\
\vec{k}_1 =&~ \vec{v}_n \times \vec{\omega} + \vec{\epsilon} \\
\vec{k}_2 =&~ (\vec{v}_n + \frac{1}{2}\vec{k}_1 {\Delta t}) \times \vec{\omega} + \vec{\epsilon} \\
\vec{k}_3 =&~ (\vec{v}_n + \frac{1}{2}\vec{k}_2 {\Delta t}) \times \vec{\omega} + \vec{\epsilon} \\
\vec{k}_4 =&~  (\vec{v}_n +\vec{k}_3\Delta t) \times \vec{\omega} + \vec{\epsilon}
\end{align}
where $\vec{\omega} =q\vec{B}/m$ and $\vec{\epsilon}=q\vec{E}/m$.
We obtain new states:
\begin{align}
\vec{v}^{n+1} =&~ \vec{v}_n + (\vec{v}_n \times \vec{\omega} + \vec{\epsilon} )\Delta t
+ \frac{1}{2}((\vec{v}_n \times \vec{\omega} + \vec{\epsilon}) \times \vec{\omega})\Delta t^2 \nonumber \\
&+ \frac{1}{3!}(((\vec{v}_n \times \vec{\omega} + \vec{\epsilon}) \times \vec{\omega}) \times \vec{\omega}) \Delta t^3 \nonumber \\
&+ \frac{1}{4!}((((\vec{v}_n \times \vec{\omega} + \vec{\epsilon}) \times \vec{\omega}) \times \vec{\omega}) \times \vec{\omega}) \Delta t^4
\\
\vec{x}^{n+1} =&~ \vec{x}_n + \vec{v}_n \Delta t
+ \frac{1}{2}(\vec{v}_n \times \vec{\omega} + \vec{\epsilon})\Delta t^2 \nonumber \\
&+ \frac{1}{3!}((\vec{v}_n \times \vec{\omega} + \vec{\epsilon}) \times \vec{\omega} )\Delta t^3 \nonumber \\
&+ \frac{1}{4!}(((\vec{v}_n \times \vec{\omega} + \vec{\epsilon}) \times \vec{\omega}) \times \vec{\omega}) \Delta t^4
\end{align}
In a uniform field,
the Jacobian for the Runge--Kutta method is
\begin{align}
J_{\rm RK4}
=
\Big| \frac{\partial(\vec{x}^{n+1},\vec{v}^{n+1})}{\partial(\vec{x}^{n},\vec{v}^{n})} \Big|
=
\begin{vmatrix}
I &
\cfrac{\partial \vec{x}^{n+1}}{\partial \vec{v}^{n}} \\
0 &
\cfrac{\partial \vec{v}^{n+1}}{\partial \vec{v}^{n}} \\
\end{vmatrix}
=
\Big| \frac{\partial \vec{v}^{n+1}}{\partial \vec{v}^n} \Big|
.
\label{eq:RK4J}
\end{align}
Without losing generality, 
we consider the case of $\vec{B}=(0,0,B)$.
Then the Jacobean $J_{\rm RK4}$ is given by
\begin{align}
\begin{vmatrix}
1
-\frac{1}{2}(\omega\Delta t)^2 
+\frac{1}{24}(\omega\Delta t)^4 
&
- (\omega \Delta t)
+ \frac{1}{6}(\omega \Delta t)^3
& 0\\
(\omega \Delta t)
- \frac{1}{6}(\omega \Delta t)^3
&
1
-\frac{1}{2}(\omega\Delta t)^2 
+\frac{1}{24}(\omega\Delta t)^4 
& 0\\
0 & 0 &1 \\
\end{vmatrix}
\nonumber \\
=
1 - \frac{1}{72}(\omega\Delta t)^6 + \frac{1}{576}(\omega\Delta t)^8
\end{align}
In the parameter range of our interest, $(\omega\Delta t)<1$,
one can see that
the phase-space volume shrinks: $J_{\rm RK4} < 1$.


\begin{thebibliography}{}

\bibitem[Hockney \& Eastwood(1981)]{hockney}
R.~W. Hockney \& J.~W. Eastwood, {\itshape Computer simulation using particles}, McGraw-Hill, New York (1981).
\bibitem[Birdsall \& Langdon(1985)]{birdsall}
C.~K. Birdsall \& A. B. Langdon, {\itshape Plasma Physics via Computer Simulation}, McGraw-Hill, New York, (1985).
\bibitem[John M. Dawson(1983)]{dawson83}
J.~M. Dawson, ``Particle simulation of plasmas,'' Rev. Mod. Phys. {\bf 55}, 403 (1983).
\bibitem[Verboncoeur(2005)]{v05}
J.~P. Verboncoeur, ``Particle simulation of plasmas: review and advances,'' \ppcf {\bf 47}, A231 (2005).
\bibitem[Boris(1970)]{boris70}
J. P. Boris, ``Relativistic Plasma Simulation --- Optimization of a Hybrid Code,'' in {\itshape Proceedings of 4th Conference on Numerical Simulation of Plasmas}, Naval Research Laboratory, Washington D. C., pp. 3--67 (1970).
\bibitem[Vay(2008)]{vay08}
J. L. Vay, ``Simulation of beams or plasmas crossing at relativistic velocity,'' \pop {\bf 15}, 056701 (2008).
\bibitem[P\'{e}tri(2017)]{petri17}
J. P\'{e}tri, ``A fully implicit numerical integration of the relativistic particle equation of motion,'' \jpp {\bf 83}, 705830206 (2017).
\bibitem[Qiang(2017)]{qiang17}
J. Qiang, ``A fast numerical integrator for relativistic charged particle tracking,'' {\itshape Nuclear Inst. and Methods in Physics Research, A} {\bf 867} 15 (2017).
\bibitem[Umeda(2018)]{umeda18}
T. Umeda, ``A three-step Boris integrator for Lorentz force equation of charged particles,'' \cpc {\bf 228}, 1 (2018).
\bibitem[Qin et al.(2013)]{qin13}
H. Qin, S. Zhang, J. Xiao, J. Liu, Y. Sun, \& W.~M. Tang, ``Why is Boris algorithm so good?'' \pop {\bf 20}, 084503 (2013).
\bibitem[Zhang et al.(2015)]{zhang15}
R. Zhang, J. Liu, J. Liu, H. Qin, Y. Wang, Y. He, \& Y. Sun, ``Volume-preserving algorithm for secular relativistic dynamics of charged particles,'' \pop {\bf 22}, 044501 (2015).
\bibitem[Higuera \& Cary(2017)]{HC17}
A. V. Higuera, \& J. R. Cary, ``A fully implicit numerical integration of the relativistic particle equation of motion,'' \pop {\bf 24}, 052104 (2017).
\bibitem[Ripperda et al.(2018)]{ripperda18}
B. Ripperda, F. Bacchini, J. Teunissen, C. Xia, O. Porth, L. Sironi, G. Lapenta, \& R. Keppens, ``A Comprehensive Comparison of Relativistic Particle Integrators,'' \apjs {\bf 235}, 21 (2018).
\bibitem[Strang(1968)]{strang68}
G. Strang, ``On the construction and comparison of difference schemes,'' SIAM J. Numer. Anal. {\bf 5}, 506 (1968).
\bibitem[LaVeque(2002)]{laveque02}
R. J. LeVeque, ``Finite volume methods for hyperbolic problems,'' Cambridge university press, doi:10.1017/CBO9780511791253 (2002).
\bibitem[Landau \& Lifshitz(1971)]{book2} L. D. {Landau} and E. M. {Lifshitz}, {\itshape The classical theory of fields}, Pergamon press, Oxford, 3rd ed., pp. 58--59 (1975).
\bibitem[Friedman \& Semon(2005)]{friedman05}
Y. Friedman \& M. D. Semon, ``Relativistic acceleration of charged particles in uniform and mutually perpendicular electric and magnetic fields as viewed in the laboratory frame,'' \pre {\bf 72}, 026603 (2005).
\bibitem[Zenitani(2018)]{zeni18}
S. Zenitani, ``Dissipation in relativistic pair-plasma reconnection: revisited,'' \ppcf {\bf 60}, 014028 (2018).
\bibitem[Parker \& Birdsall(1991)]{parker91}
S. E. Parker \& C. K. Birdsall, ``Numerical Error in Electron Orbits with Large $\omega_{ce} \Delta t$,'' \jcp {\bf 97}, 91 (1991).
\bibitem[Wu et al.(2003)]{wu03}
Y. K. Wu, E. Forest, and D. S. Robin, ``Explicit symplectic integrator for s-dependent static magnetic field,'' \pre {\bf 68}, 046502 (2003).
\bibitem[Shadwick et al.(2014)]{shadwick14}
B. A. Shadwick, A. B. Stamm, and E. G. Evstatiev, ``Variational formulation of macro-particle plasma simulation algorithms,'' \pop 21, 055708 (2014).
\bibitem[Zhang et al.(2018)]{zhang18}
R. Zhang, Y. Wang, Y. He, J. Xiao, J. Liu, H. Qin, and Y. Tang, ``Explicit symplectic algorithms based on generating functions for relativistic charged particle dynamics in time-dependent electromagnetic field,'' \pop {\bf 25}, 022117 (2018).
\bibitem[Wolski and Herrod(2018)]{wolski18}
A. Wolski and A. T. Herrod, ``Explicit symplectic integrator for particle tracking in s-dependent static electric and magnetic fields with curved reference trajectory,'' {\itshape Phys. Rev. Accel. Beams} {\bf 21}, 084001 (2018).
\bibitem[Channell and Scovel(1990)]{channel90}
P. J. Channell and C. Scovel, ``Symplectic integration of Hamiltonian systems,'' {\itshape Nonlinearity} {\bf 3}, 231 (1990).
\bibitem[Yoshida(1993)]{yoshida93}
H. Yoshida, ``Recent Progress in the Theory and Application of Symplectic Integrators,'' {\itshape Celest. Mech. Dyn. Astron.} {\bf 56}, 27 (1993).
\bibitem[Marsden and West(2001)]{marsden01}
J. E. Marseden and M. West, ``Discrete mechanics and variational integrators,'' {\itshape Acta Numerica} {\bf 10}, 357 (2001).

\bibitem[Chin(2008)]{chin08}
S. A. Chin, ``Symplectic and energy-conserving algorithms for solving magnetic field trajectories,'' \pre {\bf 77}, 066401 (2008).

\end{thebibliography}
\end{document}